\documentclass{aastex}
\usepackage{spr-astr-addons}
\usepackage{url}

\RequirePackage{color}

\begin{document}

\title{Old stellar systems in UV: resolved and integrated properties}
\shorttitle{GCs in UV}
\shortauthors{Dalessandro E.}

\author{E. Dalessandro\altaffilmark{1}} 
\affil{Dipartimento di Fisica e Astronomia, Universit\`a degli Studi
di Bologna, Viale Berti Pichat 6/2, I--40127 Bologna, Italy}

\begin{abstract}
The UV properties of old stellar populations have been subject of intense scrutiny from the late 
sixties, when the UV-upturn in early type galaxies was first discovered.  Because of their proximity 
and relative simplicity,
Galactic globular clusters (GGCs) are ideal local templates to understand how the integrated 
UV light is driven by hot stellar populations, primarily horizontal branch stars and their progeny. 
Our understanding of such stars is still plagued by theoretical uncertainties, which are partly due to the
absence of an accurate, comprehensive, statistically representative homogeneous data-set.
To move a step forward on this subject, we have combined the {\it HST} and {\it GALEX} capabilities 
and collected the largest data-base ever obtained for GGCs in UV. This data-base is best suited 
to provide insights on the HB
second parameter problem and on the first stages of GCs formation and chemical evolution and to understand how
they are linked to the observed properties of extragalactic systems.
  
\end{abstract}


\section{Introduction}

It is fair to say that the last frontier of our growing understanding of the physics of old
stellar populations resides in the ultraviolet (UV). The behavior of old stellar populations in
the UV has puzzled astronomers for almost four decades now, and in spite of major recent
progress, there are still important gaps in our understanding of the nature of stars that
dominate the integrated light of old stellar populations in the UV, particularly the far-UV 
(FUV, Ferraro et al. 1998; O' Connell 1999, Moehler 2001; Catelan 2009; Dalessandro et al. 2011,
2012, 2013a; Schiavon et al. 2012). \\
In distant extragalactic systems one can ordinarily observe only the
integrated light of unresolved stellar populations, from which the
hope is to gain knowledge about the underlying stellar population. Galactic
globular clusters (GGCs)
play an important role in understanding the integrated UV colors of
extragalactic systems, especially the so called "UV-upturn" observed in the spectral energy 
distributions of elliptical galaxies (Code \& Welch 1979; de Boer 1982; Bertola et al. 1982;
Greggio \& Renzini 1990; O'Connell 1999).
First of all,
GCs are the closest example in nature to a single stellar
population: a system of coeval stars with similar chemical
composition.  Moreover GGCs span a large range of metallicities,
a small range of ages, and perhaps some range of helium
abundance. Hence they can be used to test the stellar evolution theory, which 
in turn is one of the basic ingredients
of the models used to interpret the integrated light of distant galaxies.
GGCs are relatively nearby objects (more than
$\sim90\%$ are located at distances $r<30$\,kpc), so their populations
can be easily resolved.  With typically more than 100,000 stars, even
relatively short-lived evolutionary stages are sampled. \\
In general the main contributors to the UV emission from any
stellar system are the hottest stars. Indeed, blue horizontal branch (HB) stars and their progeny 
are well known to
be among the hottest stellar populations in GCs and contribute substantially
to the UV radiation observed from old stellar systems (Welch \& Code 1972). 
The hottest HB
stars (extreme HB, EHB) have such a small envelope mass that most of
their post-He-core burning phase takes place at high effective
temperature ($T_{\rm eff}$), during the so called "AGB-manqu\'e phase", and
these stars never return to the asymptotic giant branch (AGB).
Another group of UV-bright stars is that of post-early AGB stars,
which after a brief return to the AGB, spend the bulk of their helium
shell burning phase at high $T_{\rm eff}$. In systems with only red HB a small floor level
of $FUV$ is provided by post-AGB stars, which  
evolve to the AGB phase with an higher envelope mass where they undergo 
thermal pulses and eventually lose their envelopes moving at higher temperatures 
at constant luminosity. \\
The relative contributions of the various types of stars and
the factors that might lead to larger or smaller populations of
UV-bright stars have remained an open question (Greggio and Renzini 1990;
Dorman et al. 1995; Lee et al.  2002; Rich et al. 2005; Sohn et al. 2006). The capability 
to predict the relative contribution of various UV emitters in GCs is strongly linked to 
our knowledge of the physical mechanisms shaping the HB morphology.
The scientific community agrees from nearly 50 years about the fact that the principal parameter
governing the shape of HBs in GCs is metallicity. The general rule is that metal-rich systems have red
HBs, while in the metal-poor ones stars are distributed on average at higher effective temperatures. 
However several exceptions to this general trend are observed: remarkable cases are those of NGC~6388
and NGC~6441 (Rich et al. 1997), which despite their
metallicity ([Fe/H]$\sim-0.6$) show some of the bluest HBs known in GGCs 
(Busso et al. 2007; Dalessandro et al. 2008). Moreover several clusters, sharing similar metal content,
reveal different HB morphologies, typical cases being the pairs NGC~5927 - NGC6388 at high metallicities
([Fe/H]$\sim-0.4$), M~3 - M~13 at intermediate-metallicity regime ([Fe/H]$\sim-1.5$) and M~15 - M~92
at low-metallicities ([Fe/H]$\sim-2.3$). These noticeable exceptions have required 
the introduction of a 
second (Freeman \& Norris 1981) and possibly a third parameter in order to explain the HB distributions
in all GGCs. 
What we can call now the \emph{i-th parameter problem} 
is still a hot topic, as stressed by several authors, we recall the reader to Catelan 2009 for a nice
review (see also Dotter et al. 2010 and Gratton et al. 2010).
Despite the huge efforts made to address this problem, its solution 
is not obvious and still different
scenarios are proposed. One of the reasons that complicates the identification of
the mechanisms -- other than metallicity -- at work in shaping the 
observed luminosity and effective temperature distribution of stars along the HB 
is that there are many possible culprits (mass-loss, age, helium abundance ...; 
see Rood 1973 for example) 
and some of them are not well constrained from theory.
Age has been identified as the natural global second parameter by many authors in the past years 
(Lee et al. 1988, 1990; Lee, Demarque \& Zinn 1994; Sarajedini \& King 1989; Dotter et al. 2010; Gratton et al.
2010).  According to this
interpretation older clusters tend to have bluer HBs, while younger ones should have on average redder
HB morphologies.
\begin{figure}[t]
\begin{center}
\includegraphics[scale=0.4]{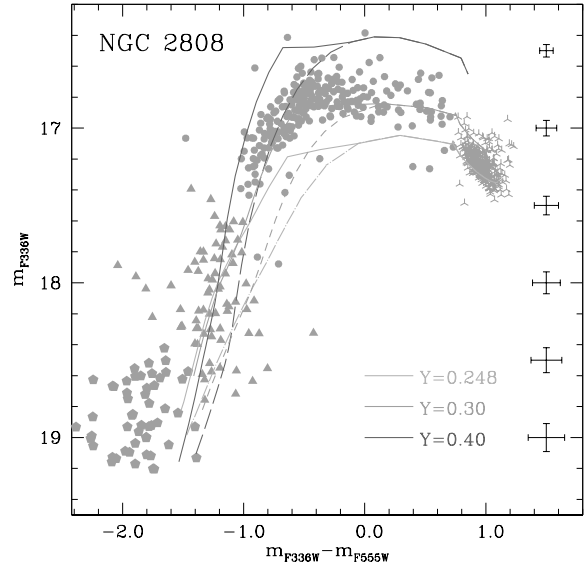}
\caption{Optical CMD of the HB of NGC~2808 compared to theoretical ZAHBs for $Y=0.248, 0.30$ and 0.40. The
dashed lines show the effect of neglecting radiative levitation in the ZAHB bolometric corrections when $T_{\rm
eff}$ is larger than 12000 K. } 
\end{center}
\end{figure}
This scenario appeared in agreement with the picture for the Galaxy formation and its early evolution
(Searle \& Zinn 1978; Zinn 1985). 
Still, age is not able to explain exhaustively the HB morphology. 
Detailed cluster to cluster comparisons 
have shown that there are systems with similar iron content and age, but 
remarkably different HB morphologies. 
The necessity of at least a third parameter transpires also from Dotter et al. (2010) and 
Gratton et al. (2010) analyses, 
in the form of either the luminosity cluster density or stellar density ($\log(\rho)$) 
-- as already suggested by Fusi Pecci et al. (1993) -- 
which might correlate with the hot extension of the HBs, or 
a variation of the initial helium abundance (Y), respectively.\\
GGCs are therefore crucial local templates for comparison with integrated
properties of distant extragalactic systems. Comparing
features in the color-magnitude diagrams (CMDs) of well known and
resolved GGCs with integrated quantities can lend important {\it model
independent} insights into the nature of extragalactic systems.

\section{A combined photometric approach}

\begin{figure}[t]
\begin{center}
\includegraphics[scale=0.4]{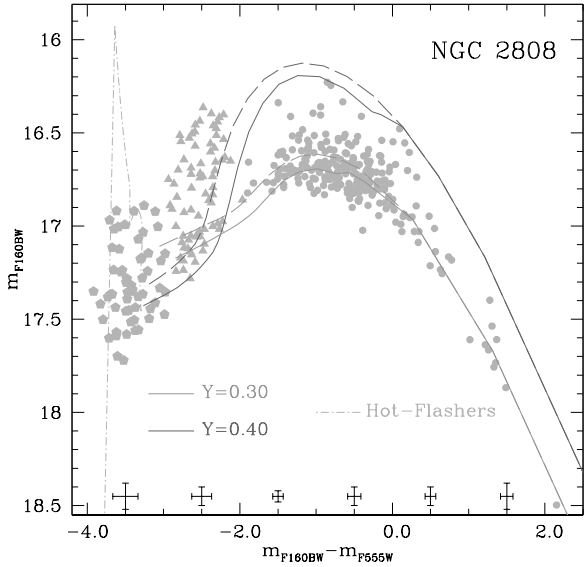}
\caption{$(m_{F160BW}$;$m_{F160BW}$-$m_{F555W}$) CMD of NGC~2808. The dash-dotted line represents the hot-flasher
model used to reproduce the BHk stars.} 
\end{center}
\end{figure}

In order to tackle this scientific subject, we used a proper combination of two complementary instruments:
the Hubble Space Telescope (HST) and the Galaxy Evolution Explorer (GALEX). HST secures high 
angular resolution ($\sim 0.05-0.1\arcsec pixel^{-1}$) and excellent photometric 
performances to resolve and study with great 
accuracy hot stellar populations even in the extremely dense cores of GCs. 
Conversely GALEX offers
a large field of view ($\sim 1deg^2$) with a much poorer spatial resolution 
($\sim 4\arcsec-5\arcsec$), which makes it suitable to sample GCs in the external regions
and to compute integrated photometry.\\
We collected observations with the Wide Field Planetary Camera 2 (WFPC2) aboard HST for more 
than 40 Galactic GCs (Prop ID: 5903 - 6607 - 10524 - 11975; PI: Ferraro) by typically using the F160BW, 
F170W, F225W, F336W and F555W
filters  (Ferraro et al. 1997, 1998, 2003, 2012; Contreras et al. 2012;
Dalessandro et al. 2008, 2009, 2013b; Lanzoni et al. 2007; Sanna et al. 2012, 2014).\\ 
We complemented this dataset with GALEX observations for 44 GCs: 38 were obtained 
as part of the GI1 and GI4 GALEX programs (PI: Schiavon) and additional 6 GGCs 
as part of GI3 (PI: Sohn). With only one exception, images were
obtained in both FUV and NUV bands. Thank to the wide FOV, it has been possible to sample the
full radial extension of basically all GCs (Dalessandro 2009, 2012, 2013b; Schiavon et al. 2012).\\
This combined database is the largest ever collected for GGCs in UV so far. It represents a 
crucial opportunity to shade new light, from one side, on the mechanisms driving 
the frequency and the temperature distribution of hot stars in GGCs and on the other to understand how 
they are linked to the integrated properties of unresolved systems.

\section{Resolved stellar photometry: the Horizontal Branch in the UV}

\subsection{The case of NGC~2808}

\begin{figure}[t]
\begin{center}
\includegraphics[scale=0.4]{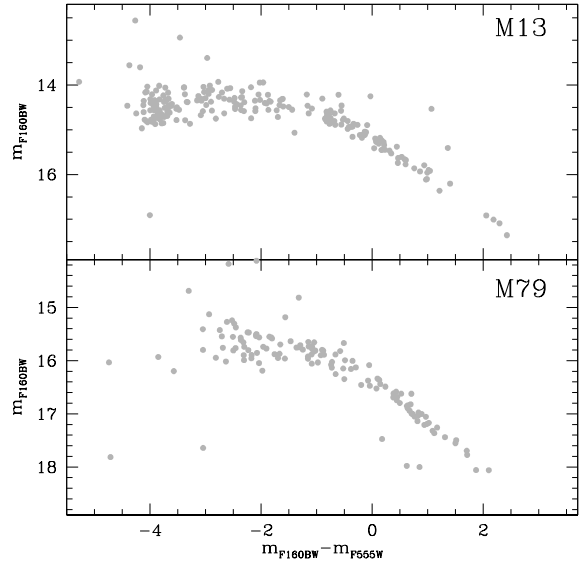}
\caption{Far-UV CMDs of M13 and M79.} 
\end{center}
\end{figure}

There is a general consensus that the first parameter shaping the globular cluster HB morphology is
metallicity and the second one has been suggested by several authors (Lee et al. 2002; Gratton et al.
 2010; Dotter et al. 2010) to be the age.
While in general this picture may be surely adequate, however it is able to explain only few
young clusters in the Galaxy, while it fails to reproduce a number of \emph{classical} cases 
where an additional ingredient (sometime called third parameter) is required.
The \emph{third parameter} has been  proposed to be 
the stellar density or luminosity density (Fusi Pecci et al. 1993; Dotter et al. 2010) or 
the initial helium abundance (Gratton et al. 2010).\\
In Dalessandro et al. (2011; see also D' Antona et al. 2005) we have shown that the main parameter 
that determines the HB morphology of NGC~2808 is Y. 
We compared high-quality UV CMDs obtained with the WFPC2 with
a large grid of suitable theoretical models (Pietrinferni et al. 2006) for which we have properly taken into account the
effect of radiative levitation. Starting from these models we produced hundreds of synthetic
HBs, which, for a given Y, are fully described by four free parameters: extinction, distance modulus, mean value of
the mass lost along the RGB ($\Delta M$) and the dispersion around the mean value ($\sigma_{\Delta
M}$). The simulated HB best reproducing the observed color distribution 
is selected on the basis of a $\chi^2$-test. 
By using this approach we have been able to satisfactory reproduce 
the complex HB morphology of NGC~2808 (see Figures~1 and 2) by assuming three different sub-populations with He abundances
compatible with what inferred from the multimodal main sequence (Piotto et al. 2007) and spectroscopic
analyses (Bragaglia et al. 2010; Pasquini et al. 2011) plus a subpopulation of "hot-flashers".  
We stress that for this kind of analyses the use of UV photometry has a crucial impact (see Figure~2). In fact 
they cannot be performed by using optical CMDs because HB sequences with different initial Y overlap
for $T_{\rm eff}> 10000$K, where the HB becomes almost vertical because of the large increase 
of bolometric corrections with $T_{\rm eff}$ (Figure~1). \\

\begin{figure}[t]
\begin{center}
\includegraphics[scale=0.4]{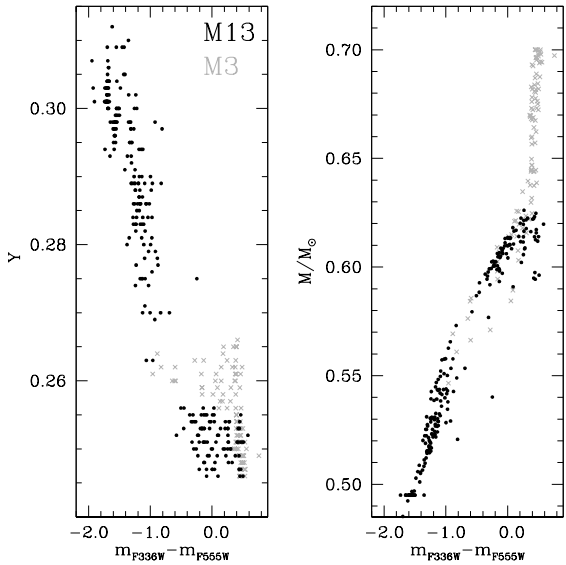}
\caption{Mass and Y distribution as a function of the ($m_{F336W}-m_{F555W}$) color along the HB of M3 (grey crosses) 
and M13 (black dots).} 
\end{center}
\end{figure}

\subsection{The classical triplet: M~3 - M~13 - M~79}
In order to further investigate the \emph{i-th parameter problem}, we selected (Dalessandro et al. 2013a)
a triplet of GGCs with
similar metallicity ([Fe/H]$\sim -1.50$) and age, namely M~3, M~13 and M~79 as templates for a comparative
analysis. Their UV CMDs obtained with the WFPC2 are shown in Figure~3\footnote{Note that for M~3
there are nor F160BW or F170W data available, therefore we have been forced to use the F225W band.}. 
This triplet is well known to have extremely different HBs, 
with M~13 displaying by far the bluest morphology and showing also evidence of at least two gaps at different temperature (Ferraro et al. 1998),
while M~3 has been proposed over the years as prototype of a "normal" HB. \\
At odds with the case of NGC~2808, these three GCs do not show 
(so far) evidences of quantized MS, so our analysis has been performed without any
a priori knowledge of the Y distribution. This forced us 
to enlarge the parameters space: the quantities needed are the minimum value of Y ($Y_{min}$) and its 
full range (${\rm \Delta Y}$), $\Delta M$  and $\sigma_{\Delta M}$. As a consequence, it is hard in these cases to determine uniquely 
a full and detailed
representation of the Y variations along the observed HBs. A more stringent and robust derivation involves 
instead the largest value of Y (${\rm Y_{max}}$) that is clearly constrained
by the distribution of stars in the $(m_{F160BW};m_{F160BW}-m_{F555W})$ CMDs.\\
We find differences  ${\rm \Delta Y_{max}}\sim0.02-0.04$ between these GGCs. In particular M~13 displays 
the largest value (${\rm Y_{max}}\sim 0.30$), M~3 (${\rm Y_{max}}\sim 0.27$) the smallest one, and M~79 is an
intermediate case with ${\rm Y_{max}}\sim 0.28$.  They seem to qualitatively correlate with  
the differences in the temperature (color) extensions of the cluster HBs. 
In Figures 4 and 5 the derived Y and mass distributions for M~3, M~13 and M~79 are shown as a function of colors. 
It is also interesting to note that our estimates of ${\rm Y_{max}}$ for these three clusters nicely
correlate with the observed range of light-element variations. In particular, M~13 shows the most extreme 
Na-O anti-correlation (Sneden et al. 2004), with stars reaching [O/Fe]$=-1.1$ and [Na/Fe]$=0.7$, while 
M~3 the least extended one, [O/Fe]$=-0.2$ and [Na/Fe]$=0.5$. This is in line with the strict
correspondence between HB colors and Na-O abundances observed in NGC~2808  by Gratton et al. (2011).\\
The comparison between M~13 and M~3 is particularly interesting, because of several previous analyses of 
their HB morphology and Y distribution. For example Catelan (2009) performed a detailed analysis 
comparing the mean masses 
of the HBs of these clusters with age differences proposed by different authors. By using several 
mass-loss recipes, he found that there is no way to reproduce the different HBs of M~3 and M~13 only in terms of age,
but at least one additional ingredient should be invoked to account for the blue HB extension observed in M~13.
We have shown (Dalessandro et al. 2013a) that M~3 and M~13 (as well as M~79) are coeval within $\sim1$Gyr.
Our best-fit model for M~13 requires to split the simulation in three steps, given that no single continuous
distribution of Y and mass loss allows to match the observed CMDs. Thus we divided the HB in  
three groups: 1) the RHB with a typical ${\rm Y_{min}}=0.246$ and a mean mass lost = ${\rm \Delta M=0.21
M_{\odot}}$, 2) stars with $-3<(m_{F160BW}-m_{F555W})<-1.5$ (Figure~3) best reproduced with ${\rm <Y>}=0.285$ and ${\rm \Delta
M=0.235M_{\odot}}$ and 3) stars with $(m_{F160BW}-m_{F555W})<-3$ which show a ${\rm <Y>}=0.30$ and ${\rm \Delta
M=0.266M_{\odot}}$. A fit with a population of stars with uniform $Y=0.265$ cannot be ruled out on the basis 
of only the HB analysis, while a single population of stars with $Y=0.28$, as proposed by Caloi \& D'Antona (2005) and 
D'Antona \& Caloi (2008), is incompatible with the distribution of stars in our UV CMD.

\begin{figure}[t]
\begin{center}
\includegraphics[scale=0.4]{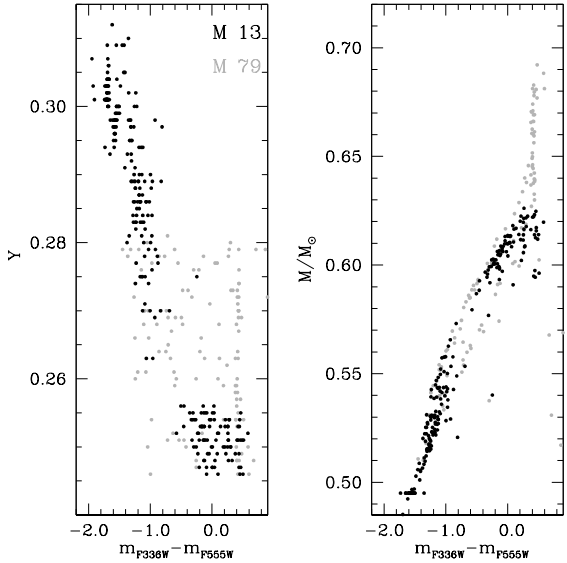}
\caption{As in Figure~4, but for M~13 and M~79.} 
\end{center}
\end{figure}

For M~3 a single synthetic population with ${\rm Y_{min}}=0.246$ and distributed according to a uniform
probability distribution with ${\rm \Delta Y}=0.02$ is required. A total mass loss of ${\rm \Delta M
=0.122M_{\odot}}$ and a linear increase as a function of Y, as constrained by the fit of star
counts as a function of magnitude and colors, is needed.\\
As highlighted by Gratton et al. (2010), while M~13 seems to behave as other relatively massive clusters,
M~3 appears to be peculiar and a more extended HB would have been expected in this case.
Our results would  lead to think that M~3 and M~13 experienced a different amount of
enrichment of light elements. This would be compatible with the scenario proposed by Gratton et al. (2010) 
(see also
Carretta et al. 2009a) that invokes a delayed cooling flow in the case of M~3. 
In particular the HB simulations and derived ${\rm Y}$ distributions would suggest
that M~13 is qualitatively similar to NGC~2808, and that it likely experienced a similar star
formation, while M~3 (and M~79) probably had a less complex formation history. \\
Analyses based on a suitable combination of UV to optical photometry and synthetic HB simulations
can provide not only insights on the HB second parameter problem, but in general 
they can potentially give some clues on the first stages of GCs formation and chemical evolution.
\\

\begin{figure}[t]
\begin{center}
\includegraphics[scale=0.6]{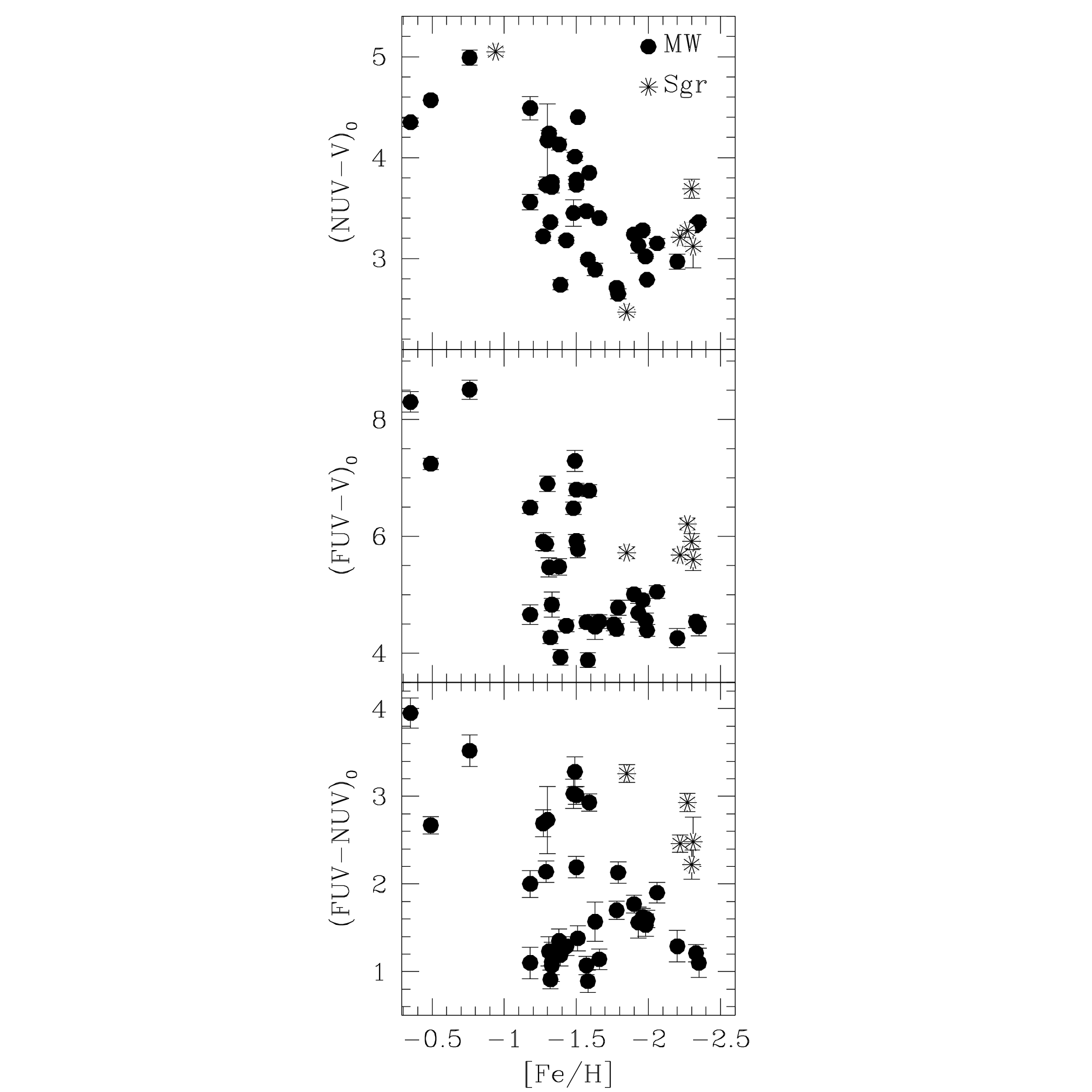}
\caption{UV integrated colors as a function of metallicity in Carretta et al. (2009b)
  scale. Clusters possibly connected with the Sagittarius stream are
  plotted as asterisks. } 
\end{center}
\end{figure}

\section{Integrated photometry: GCs properties in different environments}

GALEX FUV and NUV integrated magnitudes have been obtained by fitting the GC surface brightness
profiles (see Dalessandro et al. 2012 for details).They have been then reported
to the photometric ABMAG system by applying the zero points by Morrissey et al.
(2005).\\
As discussed in the Introduction, in old stellar populations, UV colors and in particular those involving the FUV 
are basically driven by the distribution of HB and post-HB stars. As a consequence, 
any parameter shaping the HB morphology is expected 
to drive FUV colors as well.\\
In Figure~6 we show the UV integrated colors as a function of metallicity. At a first 
glance, we observe that the three color combinations show a correlation with metallicity,
in particular,
as expected, UV color gets bluer with decreasing metallicity. However a more
careful inspection reveals more details.

\subsection{GCs in the Sagittarius Stream} 

In the (FUV-V)$_0$ and (FUV-NUV)$_0$ planes it is 
evident that the
color spread at $[Fe/H]\sim-1.5$ is due to a subset of clusters that are systematically redder
 by 1.0 - 1.5 mag in both colors
than the other GCs with the same metallicity. Interestingly these clusters (NGC~4590, NGC~5053,
NGC~5466, Arp~2 and Terzan~8) are potentially connected with the
Sagittarius dwarf galaxy stream (Dinescu et al. 1999, Palma et
al. 2002, Bellazzini et al. 2003, Law \& Majewski 2010; Carretta et al. 2014), and thus may
have an extra-Galactic origin.  This difference is qualitatively compatible with the 
extension of their HBs (Schiavon et al. 2012; Dalessandro et al. 2012). 
The Sagittarius GCs do not show any systematic trend in the
$(NUV-V)_0$ {\it vs} [Fe/H] diagram. 
We have checked that pure GGCs and Sgr clusters in our sample do not show any 
age difference, moreover
recent high resolution spectroscopic analysis (Carretta
et al. 2010) showed that, on average, the Sagittarius clusters in our
sample share the same $\alpha$-elements abundances with their Galactic
twins. 
We used the {\it R'-parameter} (Gratton et al. 2010), which is an indirect estimate of 
$Y$ (Salaris et al. 2004), to highlight possible differences.
Three (NGC~4590, NGC~5053 and NGC~5466) out of the five
Sagittarius clusters have been studied by Gratton et al. (2010). 
It is interesting to note that these clusters have {\it R'} values
smaller than other clusters with similar metallicity [Fe/H]$<-1.5$.
We performed a t-test to check the significance of the difference between
the mean values of the two distributions. It gives  a probability P$>99.9\%$ that they are different.
It emerges that clusters connected with Sagittarius share, on average, the same
properties as the genuine GGCs, except for the {\it R'-parameter}.
This difference might be an indication that those 
clusters have lower He abundances than GGCs in the same metallicity regime, and this is likely the 
main responsible of the differences in FUV integrated colors.

\begin{figure}[t]
\begin{center}
\includegraphics[scale=0.6]{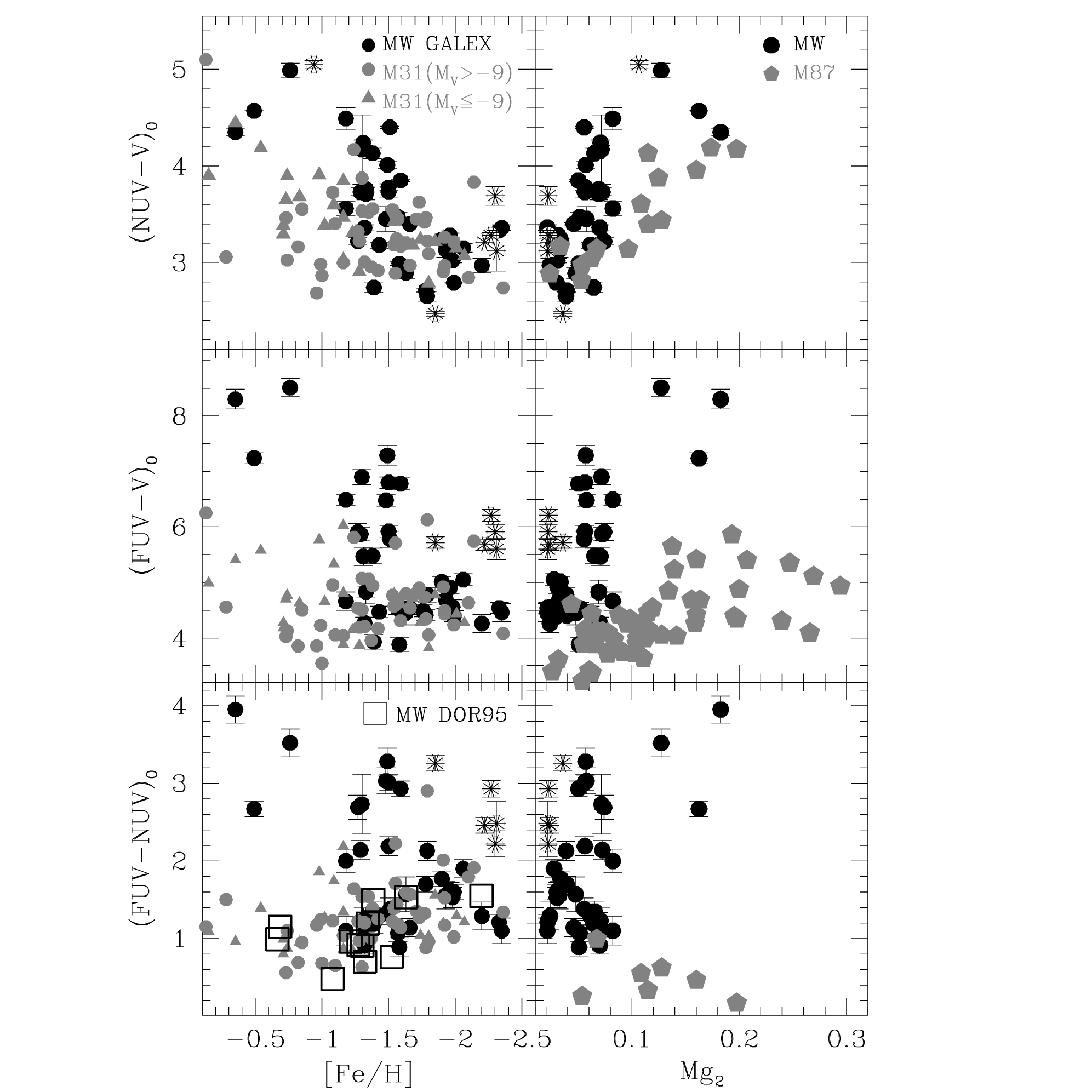}
\caption{{\it Left panels}. GALEX colors of our GGC sample (black) compared to the GCs in
  M31 (from Kang et al. 2011). The most massive M31 clusters (with
  $M_V\leq-9$) are plotted as grey triangles, while the less massive
  ones as grey circles.  In the lower panel our GGC sample has been
  supplemented with clusters observed  by Dorman et al. (1995) (open
  squares). {\it Right panels}. UV colors of GGCs (black dots and asterisks) compared to those of M87
(grey pentagons) observed by Sohn et al.  (2006), as a function of the $Mg_2$ metallicity index. } 
\end{center}
\end{figure}

\subsection{Comparison with GCs in M31 and M87}

We compared the UV colors of GGCs (left panel of Figure~7)
with those obtained by Rey et al. (2007; see
also Kang et al. 2011) for M31 clusters classified as "old" ($t>2 Gyr$; Caldwell et al. 2001) and 
we restrict the sample to clusters with  $E(B-V)<0.16$.  In the metallicity range $-2.5 <[Fe/H]< -1.0$,
the distributions in the Milky Way and
in M31 are quite similar.  The bluest colors reached
are essentially the same, and the distributions show little variations
with metallicity.  The case is very different at higher metallicity, $[Fe/H]>-1$. In the Milky Way
sample there are only red GGCs, while in M31 there are many blue
GCs. As shown in
Figure~7  roughly half of the blue, metal-rich M31 GCs are
indeed quite massive ($M_V \leq -9$).  Hence, the relative paucity of
hot, metal-rich GCs in the Milky Way could be due in part (but only in part) to
the fact that there are only two massive metal-rich clusters in our
supplemented sample (open squares). It is also possible that many GGCs with high metallicity 
and a blue HB are missed because of their location towards highly extinguished 
regions of the Galaxy.\\
We also compared GGC colors measured with GALEX with those obtained for
the giant elliptical galaxy M~87 
using HST STIS images (Sohn et al. 2006). As shown in Figure~7 (right panel), M~87 GCs are on average
bluer by $\sim1.5$ mag both in $(FUV-NUV)_0$ and $(FUV-V)_0$, while
they do not show any appreciable difference in $(NUV-V)_0$. 
On the basis of what we discussed in
Sections~3 and 4.1, we may suppose that M~87 GCs have on average a higher He content than the Milky Way
objects.  Indeed the UV color distribution of M~87 GCs can be reproduced only 
using models with enhanced values of He (Chung et al. 2011).

\section{Conclusions}

Understanding the origin and the frequency of hot stars is not simply a problem of
understanding the evolution of old, low mass stars, but it has important implications on the
interpretation of the integrated properties of galaxies. Indeed hot stars have been suggested 
to be responsible of the UV-upturn in the spectra of elliptical galaxies and bulges (Greggio \&
Renzini 1990). \\
As a part of a large project aimed at studying the properties of hot stellar populations
in GGCs as a tool to interpret the properties of unresolved stellar populations in other
galaxies, we have presented results obtained with a large UV database built with two complementary instruments:
HST and GALEX.   \\
We have shown that analyses based on a suitable combination of UV photometry
and synthetic HB simulations 
can provide not only insights on the HB second parameter problem, but in general 
they can potentially give some clues on the first stages of GCs formation and chemical
evolution (Dalessandro et al. 2011, 2013a). In fact while metallicity and age are believed to be the 
main general parameters 
shaping the HB morphology (Lee et al. 2002; Dotter et al. 2010; Gratton et al. 2010), 
there are cases that can be reproduced by only invoking different initial He abundances. In this sense
a nice example is represented by the well known triplet M~3 - M~13 - M~79. In fact these GGCs share the
same metallicity and age but have incredibly different HB morphologies. 
We find (Dalessandro et al. 2013a) differences of initial He abundances $\Delta Y_{\rm max}\sim0.02-0.04$, in
particular M~13 displays the largest value ($Y_{\rm max}\sim0.30$, M~3 the smallest one ($Y_{\rm max}\sim0.27$)
and M~79 an intermediate case  ($Y_{\rm max}\sim0.27$). These values seem to qualitatively correlate with 
the differences in temperature (color) extensions of their HBs. Interestingly they correlate also with
the observed range of light-element variations. \\
Since UV colors (FUV in particular) are primarily driven by the number of HB and post-HB stars
and their temperature distribution, we should expect He to have a quite important role on the integrated properties
of GCs. Indeed, from the comparison between GGCs and those belonging to the Sagittarius dwarf, M31 and M87, 
different behaviors
emerge. In fact the clusters associated with the Sagittarius dwarf are
 on average redder than
the MW ones, the M~31 clusters have colors which are comparable to
those of the GGCs, while the M~87 star systems are bluer. 
We note that there may be a possible trend between the mass 
of the host galaxy and
the color distribution of its globulars, in the sense that the higher is
the galaxy mass, the bluer are the GC UV colors.  In fact M~87 (with
the bluer systems) is a super-giant elliptical that is about two orders of 
magnitude more massive than the Milky Way
($1.7\times10^{13}<M/M_{\odot}<4.0\times10^{13}$; Fabricant et al. 1980), 
while Sagittarius ($\sim1.6\times10^8 M_{\odot}$; Law \& Majewski 2010) with the reddest sample of GCs
(although quite small), is a dwarf galaxy, and M31 
($3.7\times10^{11}<M/M_{\odot}<2.5\times10^{12}$; C{\^o}t{\'e} et al.2000)
and the Galaxy ($2.4\times10^{11}<M/M_{\odot}<1.2\times10^{12}$; Little \& Tremaine 1987; Kochanek 1996)
representing intermediate cases.
We argue that most
of the observed differences between colors involving the $FUV$ band
are explainable invoking different Helium contents.  This would lead
us to speculatively think that galaxies with larger masses may have, on
average, more He-rich populations. In that case, He abundance differences could be 
a by-product of chemical evolution differences, in some way connected 
to the mass of the host galaxy. This could be also connected with the 
formation and dynamical history of
clusters in galaxies with different masses, as suggested by Valcarce \& Catelan (2011; see also Carretta et al. 2014). 
In particular they argue
that clusters hosted by more massive galaxies are more likely to undergo a more complex history of
star formation thus having a larger spread in stellar populations properties.

\acknowledgments
This research is part of the project COSMIC-LAB funded by the
European Research Council (under contract ERC-2010-AdG-267675).
Photometric catalogs and integrated colors can be downloaded from http://www.cosmic-lab.eu.

\nocite{*}
\bibliographystyle{spr-mp-nameyear-cnd}
\bibliography{biblio-u1}

\end{document}